\documentclass{article}

\usepackage{fleqn,epsfig,espcrc2}

\def\bbox#1{{ \mbox{\boldmath $#1$}} }

\title{Monte Carlo studies of antiferromagnetic
spin models \\ in three dimensions}

\author{
J.~L.~Alonso, A.~Taranc\'on,\\
 \it Departamento de F\'{\i}sica Te\'orica, Facultad de Ciencias,\\
 \it Universidad de Zaragoza, 50009 Zaragoza, Spain\\
 \rm H.~G.~Ballesteros, L.~A.~Fern\'andez,
        V.~Mart\'{\i}n-Mayor and A.~Mu\~noz~Sudupe\\
 \it Departamento de F\'{\i}sica Te\'orica I,
     Facultad de Ciencias F\'{\i}sicas,\\
 \it Universidad Complutense de Madrid, 28040 Madrid, Spain\\
}

\begin{document}

\begin{abstract}
We study several antiferromagnetic formulations of the O(3) spin
model in three dimensions by means of Monte Carlo simulations. We discuss
about the vacua properties and analyze the phase transitions. Using
Finite Size Scaling analysis we conclude that all phase transitions
found are of first order.
\end{abstract}

\maketitle
\section{Introduction}

The study of antiferromagnetic models (AFM) may give us some insights
on the formulation of non trivial relativistic quantum field
theories, as they present a very rich phase space and, presumably,
new universality classes~\cite{ISING4}.
They are also interesting as related with models describing high
temperature superconductors~\cite{UNO}; the two dimensional quantum
Heisenberg antiferromagnet in the low temperature region can be
described by a non linear $\sigma$ model in three dimensions~\cite{UNO}.

We study in this contribution several formulations of three dimensional
antiferromagnetic sigma models as a first step in these directions.

It is clear that we must go beyond the naive formulation of an O(3)
$\sigma$ model in a cubic lattice with only nearest neighbors coupling
\begin{equation}
-S=\beta\sum_{\rm nn}\bbox{\sigma}_i\cdot\bbox{\sigma}_j\ ,\label{SIGMA}
\end{equation}
where $\bbox{\sigma}_i$ is a three components normalized real vector,
as it can be easily seen that the antiferromagnetic system ($\beta<0$)
is trivially related with the ferromagnetic one obtained by the
staggered transformation

\begin{equation}
\bbox{\sigma}_{x,y,z};\beta \to
\bbox{\sigma}_{x,y,z}(-1)^{x+y+z};-\beta
\label{STAGG}
\end{equation}

We will consider three models that  break
the symmetry (\ref{STAGG}) making the vacuum
frustrated:

1. A two parameter model with nearest (nn) and next to nearest
(nnn) neighbor couplings
\begin{equation}
-S=\beta_1 \sum_{\rm nn} {\bbox \sigma}_i \cdot {\bbox \sigma}_j +
   \frac{1}{2}\beta_2 \sum_{\rm nnn} {\bbox \sigma}_i \cdot {\bbox \sigma}_j\
\label{TWOPAR}
\end{equation}
Under the transformation (\ref{STAGG}) the second sum in
(\ref{TWOPAR}) does not change so it is not possible to map the
negative $\beta_2$ values onto positive ones.

2. A face centered cubic lattice (FCC) whose geometry explicitely
breaks the symmetry (\ref{STAGG}). Other cubic
lattices like the BCC (interior centered cubic) and the tetrahedrical
(diamond) fail to do so. Notice also that by setting $\beta_1=0$
in (\ref{TWOPAR}) two FCC sublattices are decoupled.

3. A Fully Frustated model constructed by defining the following set of
couplings
\begin{equation}
\begin{array}{rcl}
\beta_{x,y,z;0}&=&\beta(-1)^{x+y}\ ,\\
\beta_{x,y,z;1}&=&\beta(-1)^{z}\ ,\\
\beta_{x,y,z;2}&=&\beta\ ,
\end{array}\label{FFDEF}
\end{equation}
where $\beta_{x,y,z;\mu}$ is the coupling of the link pointing in the
$\mu$ direction from the $x,y,z$ lattice site, the values $\mu=0,1,2$
correspond to the $x,y,z$ directions respectively. This model presents
a $Z_2$ local gauge symmetry: it is invariant under a change of the
sign of a particular spin and a simultaneous change of the sign of the
couplings at the links starting from the same site.

\section{The simulation}
We have used mainly the Metropolis algorithm for the updating with
several overrelaxation steps. For the largest lattice size ($L=64$)
the number of Monte Carlo sweeps performed after thermalization, has
been of the order of $10^6$, while for the smaller lattices that
number has been even greater. We have checked in all cases that the
autocorrelation time was much smaller than the total Monte Carlo time
used for measures.

We have found the Wolff's single cluster algorithm~\cite{WOLFF}
to be very inefficient near to the antiferromagnetic transition,
because the size of the biggest cluster ussually represents a very
large fraction of the total lattice volume. For this reason, we have
used it only to study the ferromagnetic transition.

As observables we have measured the energies
\begin{eqnarray}
E_1&=&\displaystyle {\frac{1}{3V}}\sum_{\rm nn} {\bbox{\sigma}_i \cdot
\bbox{\sigma}_j}\ ,\\
E_2&=&\displaystyle {\frac{1}{6V}}\sum_{\rm nnn} {\bbox{\sigma}_i \cdot
\bbox{\sigma}_j}\ ,
\label{ENERGIAS}
\end{eqnarray}
with $V$ being the lattice volume. In the case of the FCC lattice we
only measure $E_2$, while in the Fully Frustrated model
(see~\cite{O3A} for details) the sign of the coupling has to be
properly taken into account.

For antiferromagnetic phases the standard definition of the
magnetization as
$\bbox{M}=\frac{1}{V}\sum_{i} {\bbox \sigma_i}$ is not an order
parameter. For the first two models we have instead
considered a staggered magnetization defined as
\begin{equation}
\bbox M^s_a={\displaystyle\frac{1}{V}\sum_i{(-1)}^{a}{\bbox \sigma_i}}\ ,
\label{MAGNETSTAGG}
\end{equation}
where $a=x,y,z$.
For the third model we have constructed the following set of vectors
\begin{equation}
\bbox{M}_{p}^{(i,j,k)}={\frac {8} {V}}\sum_{\stackrel{x,y,z}{{\rm (even)}}}
{\bbox \sigma}_{x+i,y+j,z+k},
\label{MAGPERIODICA}
\end{equation}
with $i,j,k=0,1$.  It can be checked that the mean values of the
previous quantities are independent of $i,j,k$. We shall refer to it
as the {\it period two magnetization}.

In practice we measure the magnetizations squared from which we
compute the Binder cumulant and the susceptibility defined
respectively as

\begin{equation}
\begin{array}{rcl}
U_L&=&1-{\langle \bbox{M}^4\rangle }/{3 {\langle \bbox{M}^2\rangle}^2}\,\\
\chi&=&V\left(\langle \bbox{M}^2\rangle -{\langle|\bbox{M}|\rangle }^2\right)\
{}.
\end{array}
\end{equation}
Another interesting quantity is the correlation length. To avoid
problems with fluctuations and asymmetric lattices we use the second
momentum definition considered in ref.~\cite{SOKAL}, valid only for
the disordered phase
\begin{equation}
\xi=\left(\frac{g_0/g_1-1}{4\sin^2(\pi/L)}\right)^{1/2}\ .
\label{XI}
\end{equation}
where $g_0$ and $g_1$ are the Fourier transforms of the propagator
\hbox{$G(\bbox{r}_i-\bbox{r}_j)=\langle \bbox{\sigma}_i \cdot\bbox{\sigma}_j
\rangle$}
at zero and minimal nonzero momentum respectively.

The derivatives of the energies and magnetizations with respect to the
couplings can be computed as connected correlations. For instance, the
specific heat matrix can be expresed as
\begin{equation}
C_{i,j}=\frac{\partial E_i}{\partial \beta_j}
        =3V\left(\langle E_i E_j\rangle-\langle E_i\rangle \langle E_j
\rangle \right)
\label{ESPECIF}
\end{equation}

We have used a Finite Size Scaling Analysis to compute the critical
point and the critical exponents associated with the phase
transitions. Measuring for instance the maxima of the specific heat
and the susceptibility, using the spectral density method, we obtain
\begin{equation}
C \sim L^{\alpha/\nu} \quad,\quad\chi \sim L^{\gamma/\nu}.
\end{equation}
The critical temperature can be obtained from the scaling behavior of
the crossing point $\beta_{L_1,L_2}$
\begin{equation}
\frac{1}{\beta_c}-\frac{1}{\beta_{L_1,L_2}}\sim
\frac{1}{\log(L_1/L_2)}.
\end{equation}
The critical exponent $\beta$ may be computed from the magnetization
\begin{equation}
\langle |\bbox{M}|\rangle_{\beta_c}  \sim L^{-\beta/\nu}\ .
\end{equation}
For first order phase transitions however the scaling behavior
presents fictitious critical exponents

\begin{equation}
\nu=\frac{1}{d}\ ,\quad \alpha=1\ ,\quad \gamma=1\ .\label{FIRSTEXP}
\end{equation}

\section{Results}

\begin{figure}[htb]
\centering\epsfig{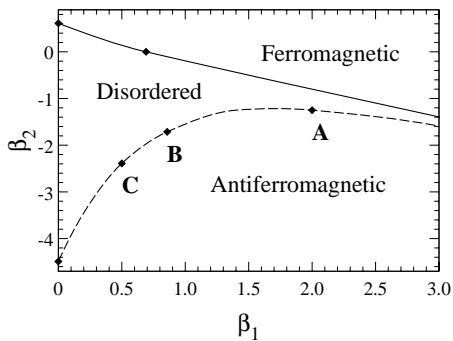}
\caption{Phase diagram for the two parameter model. The solid line
corresponds to the P-F transition and the dashed line to the P-A
transition.  The point where $\beta_2=0$ is the standard $\sigma$
model critical point. The points where $\beta_1=0$ correspond to the
critical ferromagnetic and antiferromagnetic critical points of the
FCC model.}
\label{DIAGRAMA}
\end{figure}

We have analyzed the phase diagram of the two parameter model (see
figure\ref{DIAGRAMA}). We have found three phases: ferromagnetic and
antiferromagnetic separated by a disordered (paramagnetic) phase. The
order parameter for the paramagnetic-ferromagnetic (P-F) transition
line is the magnetization, while for the
paramagnetic-antiferromagnetic (P-A) we use the staggered
magnetization.
Along the P-F line we find, for $\beta_2=0$ the standard $\sigma$
model critical point, for $\beta_1=0$ the original lattice is
decoupled into two sublattices and it is equivalent to the
ferromagnetic FCC model with $\beta=0.619(5)$. We have measured the
critical exponents on those points along the P-F line, checking that they
agree well with known values of the standard $\sigma$ model. The
exponent $\nu$ has been obtained with a 2\% of accuracy.

Along the P-A line we have measured along fixed directions the
specific heat, susceptibility, staggered magnetization and correlation
length at the following points in the $(\beta_1,\beta_2)$ plane
\begin{equation}
\begin{array}{rcl}
{\bf A}&=&(\beta_1=2,\beta_2=-1.25111(13)) \\
{\bf B}&=&(\beta_1=0.85763(8),\beta_2=-2\beta_1) \\
{\bf C}&=&(\beta_1=0.5,\beta_2=-2.3899(12))
\end{array}
\end{equation}

We have found very difficult to reach the asymptotic behavior even for
large lattice sizes. We should emphasize that it is even harder to
attain at the point {\bf B}. Assuming the results of the $L=64$ as
asymptotic we found that the growth of the specific heat is compatible
with an $\alpha/\nu=3$ while lower exponents can be readily discarded.

\begin{figure}[htb]
\centering\epsfig{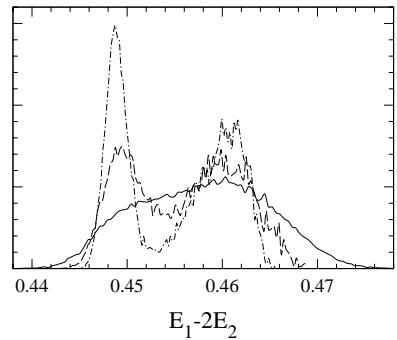}
\caption{Energy histogram for $L=32$ (solid), $L=48$
(dashes) and $L=64$ (dot-dashes) at {\bf B}.}
\label{HISTOG}
\end{figure}

{}From an analysis of the magnetic susceptibility we exclude a second
order transition. The energy histograms (see figure\ref{HISTOG}) confirm
this asumption, for large enough $L$ they present a clear two peak
structure with an stable inter-peak distance. We point out that even
for the smaller lattice sizes where the two peaks cannot be resolved
the specific heat and the susceptibility give strong indications of
the transition first order character.

To estimate the correlation length according to (\ref{XI}), we have to
simulate in the disordered phase and then extrapolate the results in
order to obtain the values at the critical point.
We have found the following values for the correlation lengths defined
in (\ref{XI})
\begin{equation}
\xi^{\bf A}\sim 7 \quad
\xi^{\bf B}\sim 12\quad
\xi^{\bf C}\sim 7\ ,
\label{XIMAX}
\end{equation}
with statistical errors of the order of 10\%. These values explain
{\it a posteriori} the difficulties found in reaching, especially for
point {\bf B}, the asymptotic region.

The second model considered is the FCC lattice with antiferromagnetic
coupling. Classically the ground state presents a O(3)$^L$ degeneracy
group. However, when thermal fluctuations are taken into account a
collinear ground state is selected (see~\cite{O3A} for details). This
is an example on Villain's {\it order from disorder}~\cite{QUINCE}. In
the absence of another interaction that fixes the global direction,
there remains a $Z_2^L$ degeneracy.

We have performed a numerical simulation in the $L=24$ lattice in the
low temperature phase ($\beta=-5<\beta_c$) that confirms the collinear
ground state structure. Each of the $2^L$ ground states is very
stable under Monte Carlo evolution with a local update algorithm.
Extrapolating to infinite volume the transition point, in the
assumption of a first order behavior, we obtain
\begin{equation}
\beta_c^{\rm FCC}=-4.491(2).
\end{equation}
The energy histogram of the $L=32$ lattice presents a clear double
peak that establishes the first order nature of the transition,
although for that lattice size the asymptotic behavior has not been
reached.

Finally let us present our results for the Fully Frustrated model. As
the hamiltonian of this model is invariant under the transformation
(\ref{STAGG}) we will consider only the $\beta\geq 0$ case. We obtain a
phase transition, between a disordered phase for small $\beta$ and
an ordered one with a complicated structure, at
\begin{equation}
\beta_c^{\rm FF}=2.26331(13)
\end{equation}

The ground state in the ordered phase is highly degenerate. We have
checked for $L=2$, by means of numerical and analytical methods, that
the equilibrium configurations, in the \{$e_1,e_2,e_3$\} space where
$e_i\equiv \bbox{\sigma}(\bbox{r}_0)\cdot\bbox{\sigma}(\bbox{r}_i)$,
lay inside the hexagon perpendicular to the (1,1,1) vector with
vertices at $(1,\frac{2\sqrt{3}-3}{3},\frac{\sqrt{3}}{3})$ and
permutations. For $L>2$ the equilibrium configurations concentrate
around the six corners of the hexagon in a region whose size decreases
with increasing lattice size.

The finite size scaling analysis of the specific heat shows that this
is a weak first order phase transition with a hard to reach
thermodynamic limit. From the $L=8$ to $L=24$ values we find good
agreement with $\alpha/\nu=1$ while from the $L=24$ to $L=64$ we find
an increasing value that tends to $\alpha/\nu=3$ as expected for a
weak first order phase transition. We only observe an incipient two
peak structure in the energy histogram for $L=64$. For the
susceptibility we find a behavior compatible with a first order
character. From the bigger lattices we quote $\gamma/\nu\sim 3$. We
therefore conclude that the order parameter for the transition,
i.e. the period two magnetization becomes a discontinuous function of
$\beta$ in the thermodynamic limit.

\section{Conclusions}

We have explored three models with internal O(3) symmetry that are
antiferromagnetic or develop frustration. We conclude that all the
antiferromagnetic transitions found are of first order.

We have checked that the ferromagnetic line belongs to what seems to
be the only  universality class for O(3) models in three dimensions.

\subsection*{Acknowledgements}

We thank to Gabriel \'Alvarez Galindo, Alan Sokal, and Arjan Van der
Sijs for useful discussions. Partially supported by CICyT AEN93-0604,
AEN94-0218 and AEN93-0776. H.~G.~Ballesteros and V.~Mart\'{\i}n-Mayor are MEC
fellows.

\end{document}